\documentclass[tighten]{PoS}
\usepackage{subcaption}
\usepackage{graphicx}
\usepackage{cite}
\usepackage[hidelinks]{hyperref}

\title{Correlations and lags between X-ray and radio emission of Cyg X-1}

\ShortTitle{Correlations and lags in Cyg X-1}

\author{\speaker{J. N. S. Shapopi}$^a$ and Andrzej A. Zdziarski$^b$\\
        \llap{$^a$}Department of Physics, University of Namibia\\ Windhoek, Namibia\\
        \llap{$^b$}Nicolaus Copernicus Astronomical Center, Polish Academy of Sciences\\ Bartycka 18, PL-00-716 Warsaw, Poland\\
        E-mail: \email{jimzashapopi@yahoo.com}, \email{aaz@camk.edu.pl}}

\abstract{Cygnus X-1 is a Galactic black-hole binary and a microquasar. Its X-ray emission originates most likely in the accretion flow, while the radio emission is from a magnetized jet. We study long-term cross-correlations between its X-ray and radio fluxes, searching for lags of the peak of the correlation. We use soft X-ray data from the All Sky Monitor on board of {\it Rossi X-ray Timing Explorer}, hard X-ray data from the Burst Alert Telescope on board of {\it Neil Gehrels Swift}, and 15 GHz radio monitoring data from the Ryle and AMI telescopes, which light curves cover all of the spectral states of Cyg X-1. Previous studies of radio/X-ray correlations in black-hole binaries concentrated on the soft X-ray/radio correlation in the hard spectral state. In contrast, our strongest correlation is between the hard X-rays and radio in the entire data at a $\lesssim 1$-day lag. This indicates the jet formation from a hard X-ray emitting corona/hot accretion flow through all states of Cyg X-1. We also find evidence for a long lag of the radio with respect to the soft X-rays in the soft state. This indicates long-term advection of the magnetic field through the accretion disc from its high-mass donor, which is similar to the case of another microquasar with a high-mass donor, Cyg X-3. }

\FullConference{7th Annual Conference on High Energy Astrophysics in Southern Africa ***\\
		28--30 August 2019\\
		Swakopmund, Namibia}

\begin{document}

\section{Introduction}

The X-ray source Cyg X-1, discovered in 1964 \cite{bowyer65}, is a high-mass binary consisting of an OB supergiant accreting via stellar wind onto a black hole \cite{gies82,gies86}. It is one of the brightest and best studied X-ray binaries, e.g.\ \cite{BH90, gierlinski97, gierlinski99, cygx1_mass}. The X-ray emission of Cyg X-1 originates most likely from its accretion flow, while the radio emission is from its jet, resolved by VLBA and VLA observations \cite{stirling01}. 

Cyg X-1 shows two main spectral states, hard \cite{gierlinski97} and soft \cite{gierlinski99}, distinguished by the shape of the X-ray spectra. In the hard state, disc emission is weak and the spectrum is dominated by a hard power law with a high energy cut-off at $\gtrsim$100 keV, well modelled by thermal Comptonization \cite{gierlinski97}. In the soft state, the blackbody disc emission \cite{ss73} dominates the spectrum, and it is followed by a high energy tail with varying strength, well fitted by coronal Comptonization \cite{gierlinski99}.

\section{The Data and Analysis Methods}

We study long term correlations between the 15 GHz radio emission and the soft and hard X-rays. We use the radio monitoring light curves for MJD 50226--53902 from the Ryle telescope, and for MJD 54573--57748 from the Arcminute Microkelvin Imager Large Array, as published in \cite{Andrzej2017}. We correlate them with the light curve for MJD 50087--55870 from the All Sky Monitor (ASM) \cite{ASM} on board of {\it Rossi X-ray Timing Explorer}, which covers three bands, 1.5--3, 3--5 and 5--12 keV. We also use the light curve for MJD 53416--57848 from the Burst Alert Telescope (BAT) \cite{BAT} on board of {\it Neil Gehrels Swift}, which is for the 15--50 keV band. We correlate the radio and X-ray light curves either separately for the different states or for all of the data. To separate the states, we follow \cite{Andrzej2017}. Since the radio/X-ray correlation has similar properties in both hard and intermediate states \cite{Andrzej2011}, we treat them jointly, and hereafter use the term `hard' for the sum of the hard and intermediate states. Some of the results presented here are also given in \cite{ZSP_2020}. 

The cross-correlation coefficients are calculated using the method of \cite{E_and_K} with some modifications described in \cite{Andrzej2018}. Specifically, we calculate the Pearson's correlation coefficient for two discrete and unevenly spaced light curves, $x_i$ and $y_j$, shifted in time by $\Delta t$,
\begin{equation}
r(\Delta t) = \frac{\sum_{i,j} [x_i - \bar{x}(\Delta t)][y_j - \bar{y}(\Delta t)]/K(\Delta t)}{\sigma_x(\Delta t) \sigma_y(\Delta t)}
\label{Pearson} 
\end{equation}
where the summation is over all pairs, $(i,j)$, corresponding to the time shift of $\Delta t\pm \delta/2$,
\begin{equation}
\Delta t - \delta/2 \le t(y_j) - t(x_i) < \Delta t + \delta/2.
\label{criterion}
\end{equation}
Here, $\bar x$, $\bar y$ and $\sigma_x$, $\sigma y$ are the light curves average values and standard deviations, respectively, which are calculated for a given $\Delta t$ for the points satisfying the criterion (\ref{criterion}) only, $K$ is the number of such pairs, and $\delta$ is the bin size. (Note a typo in equation 3 of \cite{Andrzej2018}, where $y_i$ in the numerator should be $y_j$.) The standard deviation is calculated as \cite{E_and_K} 
\begin{equation}
\sigma(\Delta t) = \frac{1}{K(\Delta t)-1} \sqrt{\sum_{i,j} \left\{\frac{[x_i - \bar{x}(\Delta t)][y_j - \bar{y}(\Delta t)]}{\sigma_x(\Delta t) \sigma_y(\Delta t)} - r(\Delta t)\right\}^2},
\end{equation}
where the summation is over the same pairs as in equation (\ref{Pearson}). In our convention, positive lags correspond to radio lagging X-rays. 

\section{Results}

\vskip -0.5cm
\begin{figure}[ht!]
	\begin{subfigure}[b]{0.8\textwidth}
\centerline{
	\includegraphics[width = 0.75\textwidth]{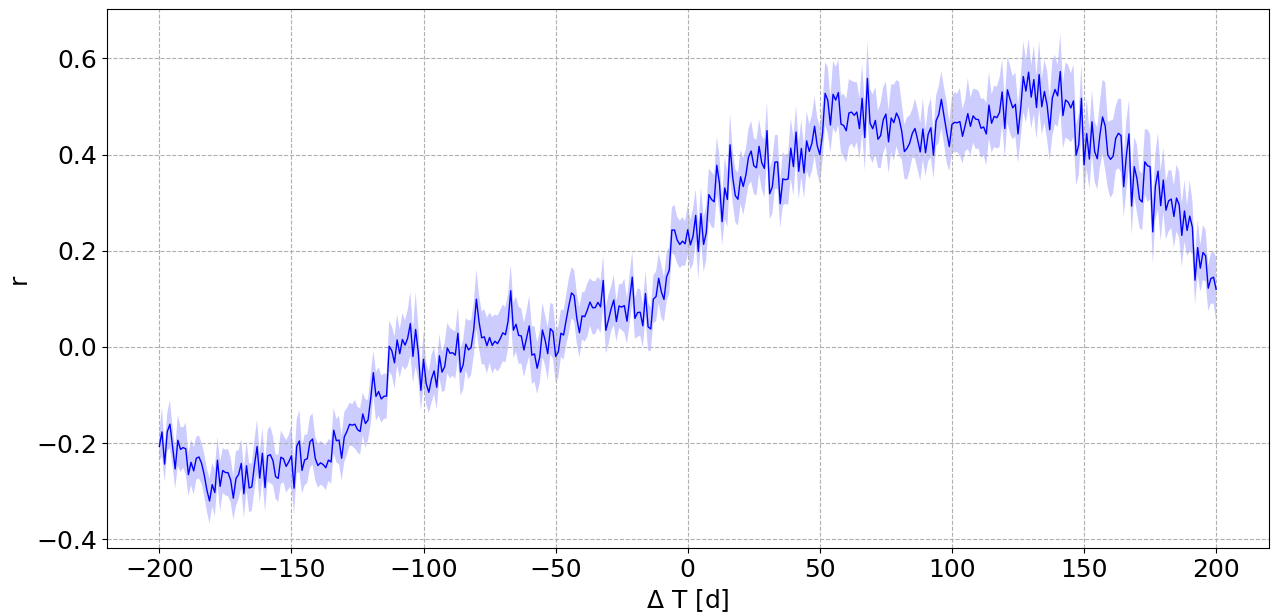}}
		\caption{15 GHz vs.\ 1.5--3 keV in the soft state.}
	\end{subfigure}
	\begin{subfigure}[b]{0.8\textwidth}
		\centering
	\includegraphics[width = 0.75\textwidth]{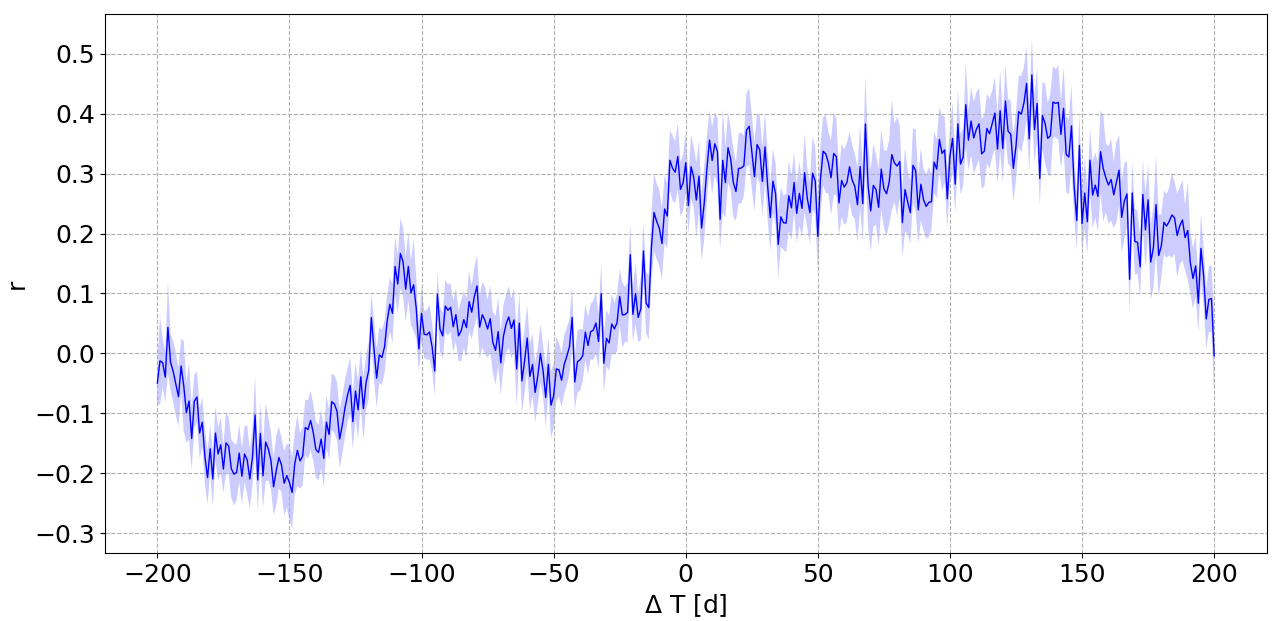}
		\caption{15 GHz vs.\ 3--5 keV in the soft state.}
	\end{subfigure}
	\begin{subfigure}[b]{0.5\textwidth}
		\centering
	\includegraphics[width=1.0\textwidth]{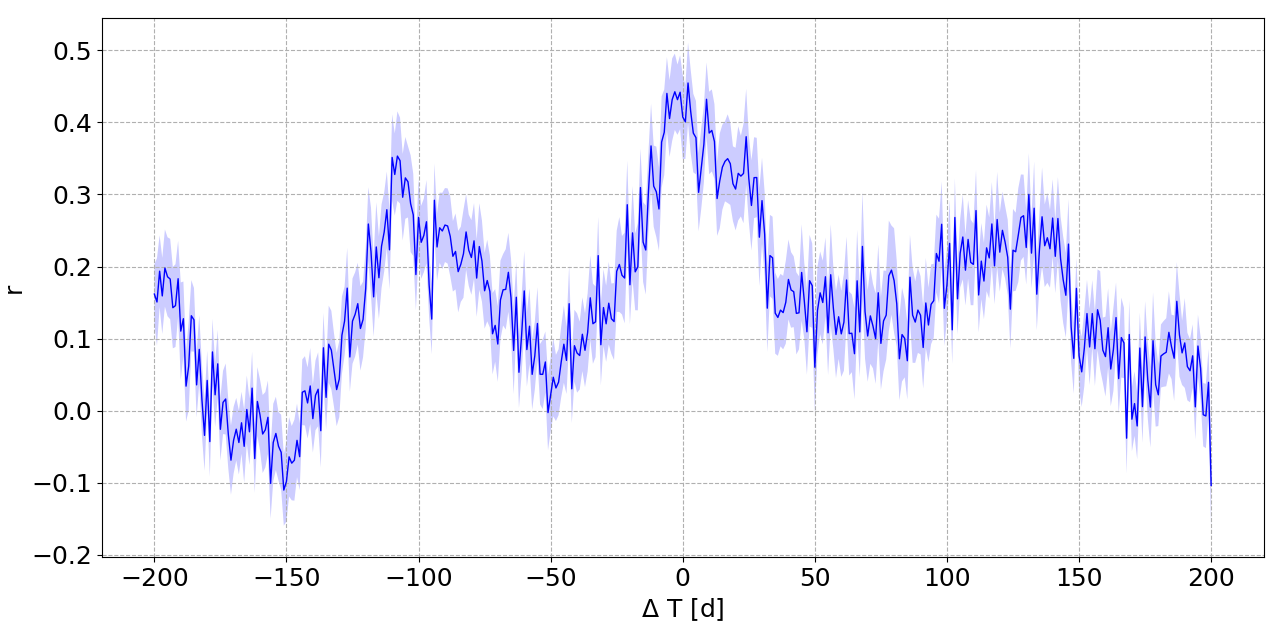}
		\caption{15 GHz vs.\ 5--12 keV (ASM) in the soft state.}
	\end{subfigure}
	\begin{subfigure}[b]{0.5\textwidth}
		\centering
		\includegraphics[width = 1\textwidth]{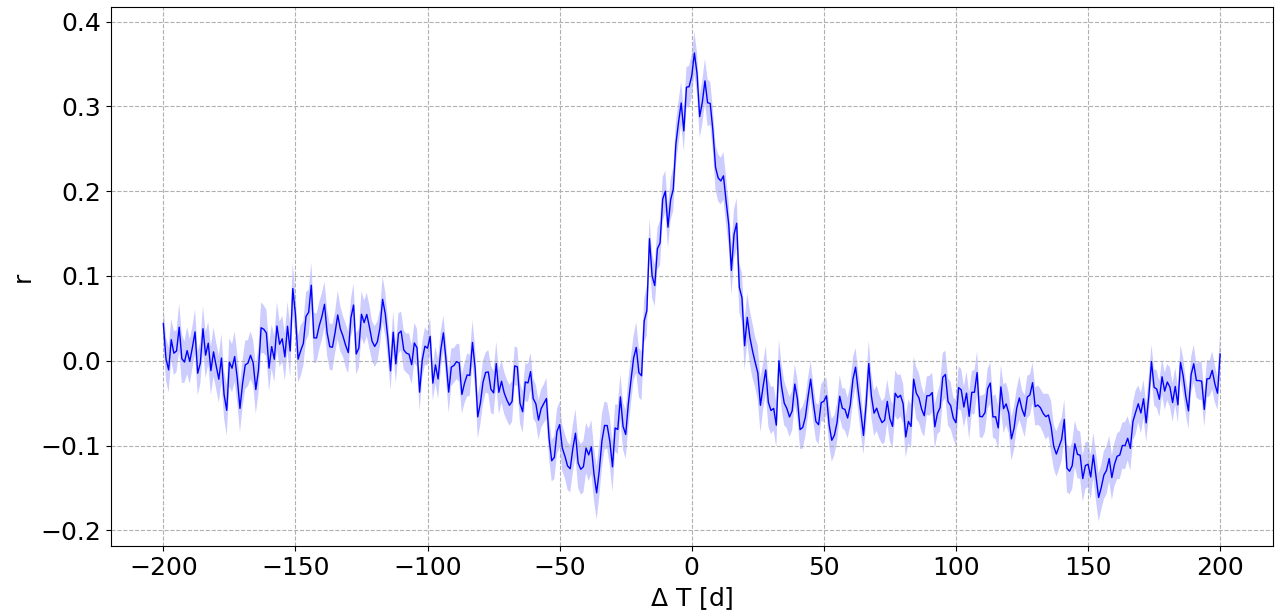}
		\caption{\small 15-GHz vs.\ 1.5--12 keV (ASM) in the hard state}
	\end{subfigure}
	\caption{ASM light curves cross-correlation with the 15 GHz radio emission. In all figures, the cross-correlations are shown by the blue curves, while the blue shades show the error ranges.}
	\label{asm}
\end{figure}

\begin{figure}[ht!]
	\begin{subfigure}[b]{0.8\textwidth}
		\centering
		\includegraphics[width = 0.75\textwidth]{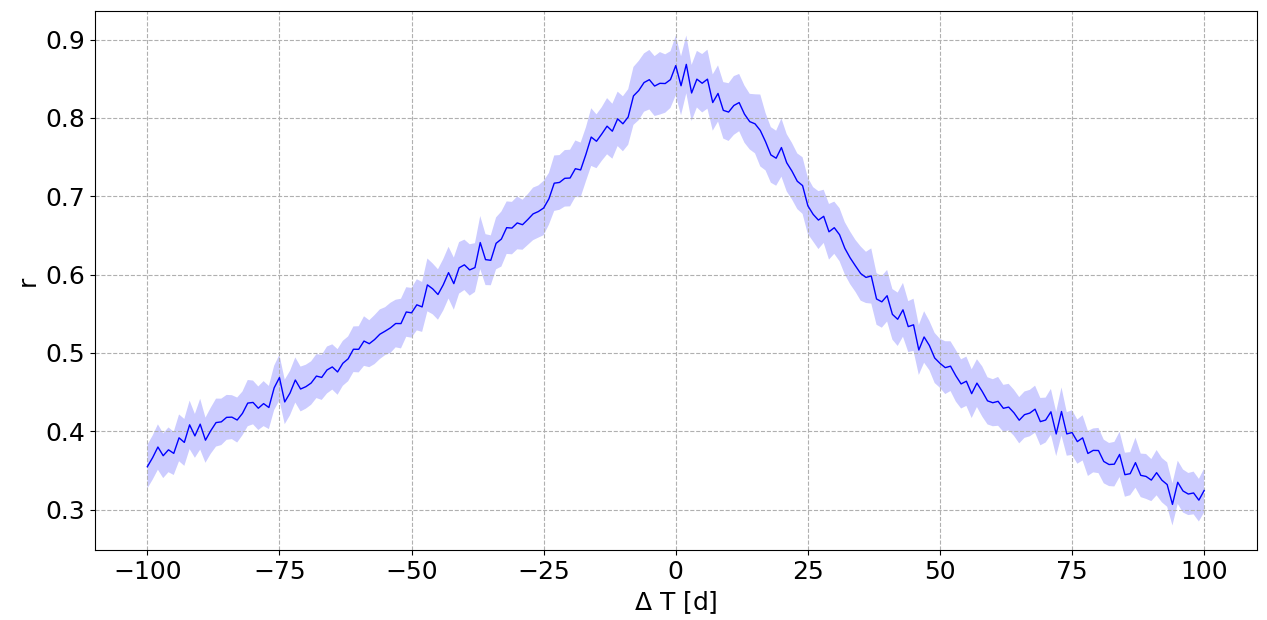}
		\caption{15-GHz vs.\ 15--50 keV (BAT) in all states.}
	\end{subfigure}
	\begin{subfigure}[b]{0.5\textwidth}
		\centering
		\includegraphics[width = 1.0\textwidth]{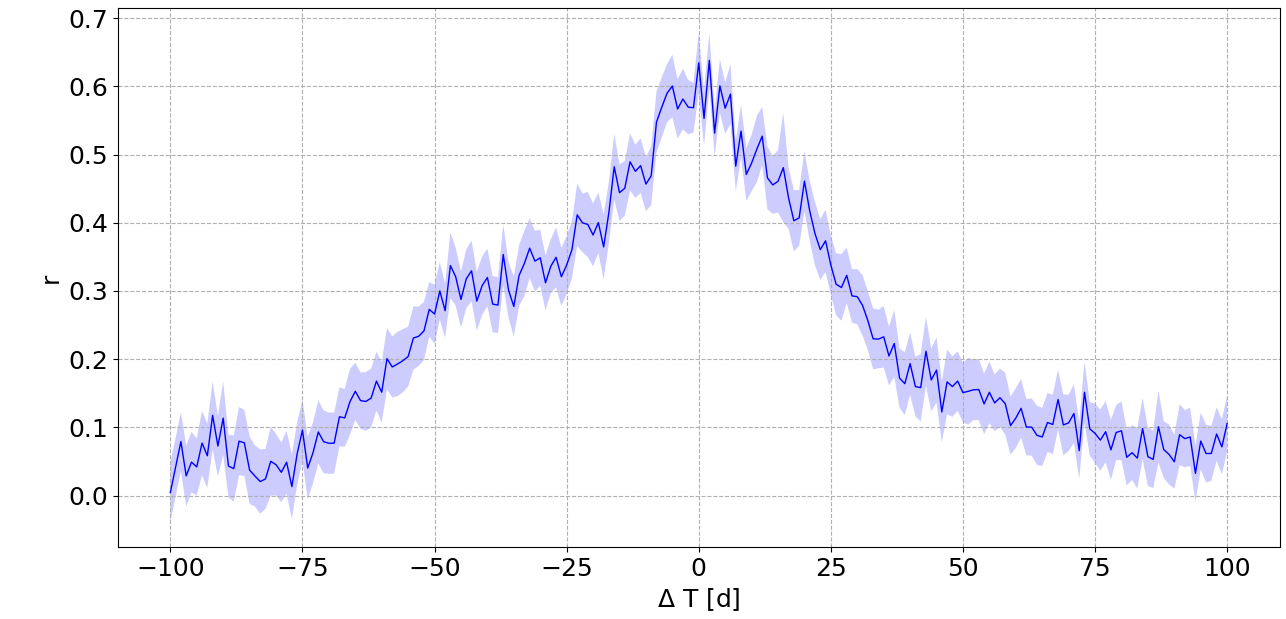}
		\caption{15-GHz vs.\ 15--50 keV (BAT) in the soft state.}
	\end{subfigure}
	\begin{subfigure}[b]{0.5\textwidth}
		\centering
		\includegraphics[width=1.0\textwidth]{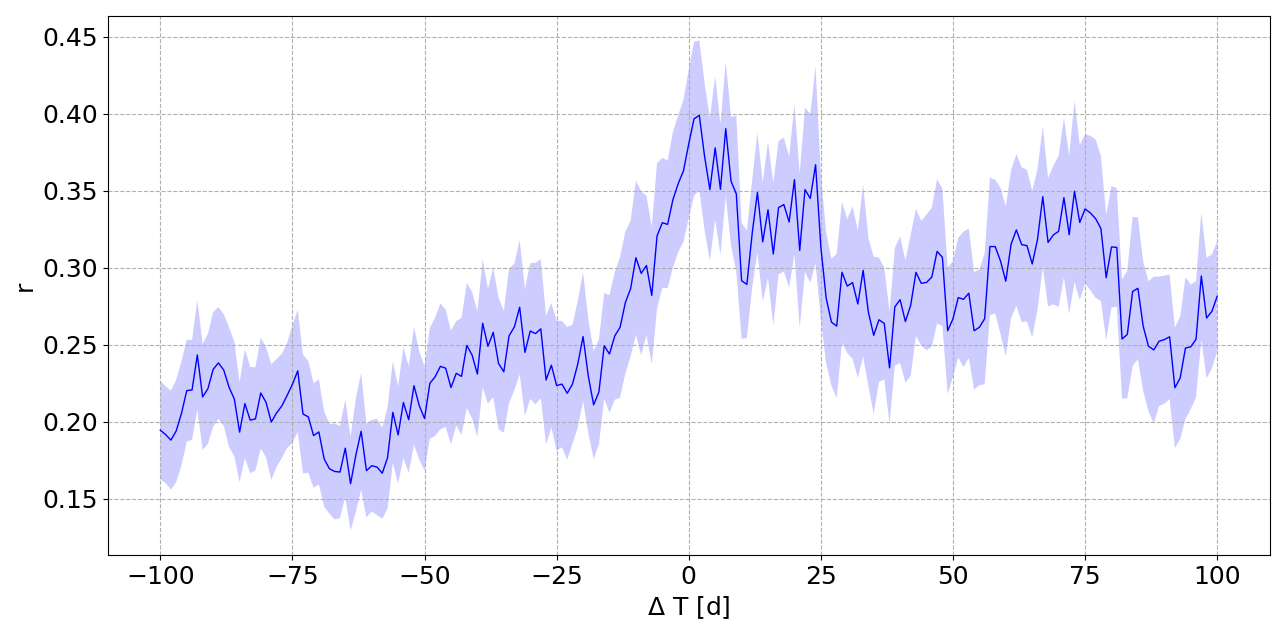}
		\caption{15-GHz vs.\ 15--50 keV (BAT) in the hard state.}
	\end{subfigure}
	\caption{BAT light curves cross-correlations with the 15 GHz radio emission.}
	\label{bat}
\end{figure}

We have calculated cross-correlations for all of the X-ray bands, and for all of the soft, hard and combined states. Most of them show significant correlations; here we show some selected, most interesting results. Fig.\ \ref{asm}(a) shows the correlation of the radio with the softest ASM band, 1.5--3 keV in the soft state. We see that while the correlation at zero lag is weak, $r\sim 0.2$, it increases with the lag up to the maximum at $\Delta t\sim 140$ d. The highest range of values of $r\sim 0.5$ occurs in the wide range of $\Delta t\sim 50$--150 d. In the 3--5 keV band and the soft state, Fig.\ \ref{asm}(b), we see a similar behaviour, but the zero-lag correlation is stronger, $r\sim 0.3$, and the maximum at a somewhat longer $\Delta t\sim 100$--150 d is weaker, up to $r\sim 0.4$. Following this trend, the 5--12 keV band in the soft state shows the global maximum around zero lag with $r\approx 0.45$, and the (local) maximum around $\Delta t\sim 100$--150 d similar to those of the lower-energy bands is now relatively weak, with $r\approx 0.2$--0.25, see Fig.\ \ref{asm}(c). For the hard state, we show only the result for the sum of the ASM bands, 1.5--12 keV. We see a distinct and relatively narrow maximum around zero lag, though with a relatively small coefficient, $r\approx 0.35$.

We then show the correlations with the BAT light curves. Fig.\ \ref{bat}(a) shows the results for all the data. We see a very high correlation coefficient, $r\approx 0.85$, around zero lag. Then, the correlation in the soft state is similar, with the maximum around zero lag with $r\approx 0.6$, see Fig.\ \ref{bat}(b). On the other hand, the correlation in the hard state is much weaker, $r\approx 0.4$ at the zero lag, and the cross-correlation profile is complex, see Fig.\ \ref{bat}(c).  

\section{Interpretation}

We have obtained two main results. Our strongest correlation found in the analysed data is that between the 15--50 keV hard X-rays and the 15 GHz radio emission around zero lag for the full light curves, i.e., including both the hard and soft states, shown in Fig.\ \ref{bat}(a). When dividing the light curve into the states, a stronger and clearer correlation is in the soft state, though weaker than that for the entire light curves.

The hard X-ray emission appears to be from Comptonization in a corona/hot accretion flow in the vicinity of the black hole \cite{gierlinski97, gierlinski99, done07}, while the radio emission is from the jets, in particular the approaching one \cite{stirling01}. Surprisingly, the correlation for the entire data is very strong, with $r$ close to unity. The correlation of radio and X-rays in the hard state, Figs.\ \ref{asm}(d), \ref{bat}(c), is similar at zero lag to that in other accreting binaries containing a black hole, e.g., \cite{corbel13}. Such correlations agree with the results of magneto-hydrodynamical simulations, showing a formation of a strong jet by a geometrically thick disc \cite{Liska20}, with the X-rays being due to thermal Comptonization in the thick hot disc \cite{gierlinski97}. However, the correlation of radio with hard X-rays in Cyg X-1 is stronger in the soft state, whereas this state in other X-ray binaries accreting onto a black hole is associated either with jet quenching or occasional ejection of individual radio-emitting blobs \cite{fender04}. In contrast, the jet in the soft state of Cyg X-1 is present almost always, and is very tightly linked to the high-energy tail beyond the disc blackbody. That tail is well modelled by Comptonization in a corona above the surface of the accretion disc \cite{gierlinski99}. This points out to the jet emission being strongly linked to the hot Comptonizing medium, either a hot disc or a hot corona, in all of the states of Cyg X-1. This conclusion was reached before by \cite{Andrzej2011}, but at a much weaker statistical significance.

Second, we have found the cross-correlation with soft X-rays in the soft state of Cyg X-1 is relatively strong and positive, but only for long lags in a wide range of $\Delta t\sim 50$--150 d. The strength of this lagged correlation is highest in the 1.5--3 keV band, as shown in Fig.\ \ref{asm}(a), and it decreases with the increasing energy. In hard X-rays, 15--50 keV, the lagged correlation disappears completely, see Fig.\ \ref{bat}(b). The cross-correlation for the 1.5--3 keV band has a very asymmetric shape, showing mostly positive lags, which appears to be highly significant statistically. 

We interpret these lags as due to the time between a fluctuation of the thin accretion disc arriving near the black hole and the subsequent response of the jet. The jet production in a thin disc requires the presence of poloidal magnetic field, e.g.\ \cite{liska19}, and this field can be advected from the donor star. A similar radio lag with respect to the soft X-rays in the soft state was found in Cyg X-3 \cite{Andrzej2018}, and the mechanism for field advection in Cyg X-1 can be similar to that proposed for Cyg X-3 by \cite{cao19}.

\acknowledgments
We thank Barbara De Marco for valuable comments. We acknowledge support from the Polish National Science Centre under the grant 2015/18/A/ST9/00746. 

\bibliographystyle{JHEP}
\bibliography{references}
\end{document}